\documentclass[twocolumn,twocolappendix]{aastex631}
\usepackage{newtxtext,newtxmath}

\begin{document}

\title{Magnetic Interactions in White Dwarf Binaries as Mechanism for Long-Period Radio Transients}

\author[0000-0003-4721-4869]{Yuanhong Qu}\thanks{E-mail: yuanhong.qu@unlv.edu}

\author[0000-0002-9725-2524]{Bing Zhang}\thanks{E-mail: bing.zhang@unlv.edu}

\affiliation{The Nevada Center for Astrophysics, University of Nevada, Las Vegas, NV 89154}
\affiliation{Department of Physics and Astronomy, University of Nevada Las Vegas, Las Vegas, NV 89154, USA}

\begin{abstract}
A growing population of long-period radio transients has been discovered and their physical origin is still up to debate. Recently, a new such source named ILT J1101 + 5521 was discovered, which is in a white dwarf (WD) -- red dwarf (RD) binary system, with the observed 125.5 min period in radio emission being identified as the orbital period and the radio emission occurs at the inferior conjunction of the WD. We suggest that the radio emission properties of the system can be well explained within the framework of the unipolar inductor magnetic interaction model between the magnetized WD and the RD with low magnetization, with a relativistic version of electron cyclotron maser emission being the most likely radiation mechanism. 
We suggest that this mechanism can interpret at least some long-period radio transients, especially the ultra-long period sub-population.
Within this model, high energy emission in X-rays via relativistically boosted cyclotron radiation and $\gamma$-rays via inverse Compton scattering off stellar light are expected, but the predicted luminosities are relatively low. This model likely applies to the ultra-long period population of LPRTs. The short-period population of LPRTs is likely powered by other engines such as slow magnetars.
\end{abstract}


\section{Introduction}

Recently, a growing population of long-period radio transients (LPRTs)\footnote{Also called ``ultra-long-period objects (ULPOs) in the literature.} have been discovered \citep{Hyman2005,Hurley-Walker2022,Hurley-Walker2023,Hurley-Walker2024,Ruiter2024}. These sources typically have a period from minutes to hours, usually with a duty cycle of the active window $\Delta t / P \ll 1$ (where $\Delta t$ is the active duration and $P$ is the period). The physical origin of these sources is unknown. The suggested engines include spindown-powered magnetized white dwarf pulsars \citep{ZhangGil2005,katz22}, ultra-long-period magnetars \citep{Wadiasingh20,Cooper2024}, and regular magnetars with long-period precession \citep{Zhu&Xu2006}. 
A related white dwarf optical pulsar source with a similar period has a red dwarf (M dwarf) companion \citep{Marsh2016,Buckley2017} and may be powered by binary interactions \citep{Geng2016}. 
Multi-wavelength observations have given constraints on some of these models \citep{Kaplan2008,Rea2022,Rea2024}, but there has been no smoking gun observational clue that helps to nail down the physical origin of these systems.

Recently, a newly discovered LPRT, named ILT J1101 + 5521, has been identified as a white dwarf (WD) -- {red dwarf (RD) system\footnote{More specifically, the companion is identified as an M dwarf, but red dwarfs in general also include K dwarfs.}}, and the measured $\sim$125.5-min period is identified as the orbital period \citep{Ruiter2024}.  
Particularly, \cite{Ruiter2024} found that the RD is behind the WD and in line with respect to Earth when the radio pulses are emitted. 
Incidentally, two other long period {optical and radio} transients, i.e. the 3.56-hr-period optical white dwarf pulsar AR Scorpii \citep{Marsh2016,Buckley2017} and the 2.9-hr-period LPRT GLEAM-X J0704-37 \citep{Hurley-Walker2024}, are also found to be associated with an RD companion. 
It seems that WD -- RD binaries likely provide the preferred physical conditions to produce pulsed emission in such systems at least for these sources.

In this paper, we propose that the coherent radio emission of ILT J1101 + 5521 is powered by a relativistic version of electron cyclotron maser emission (ECME) mechanism in the magnetic field lines connecting the WD and the RD, with electromagnetic potential along the field lines excited through the unipolar inductor effect due to the orbital motion of the weakly magnetized RD moving in the magnetosphere of the magnetized WD. 
Such a mechanism has been successfully invoked to interpret the Jupiter decametric emission due to magnetic interactions between Jupiter and Io \citep{Goldreich&Lynden-Bell1969}, and it has been speculated to also operate in other systems as well, including
ultra-compact double white dwarf binaries \citep{Wu2002,DallOsso2006},
short-orbital-period extrasolar super-earths \citep{Laine2012},
and binary neutron star or neutron star -- black hole mergers 
\citep{Hansen&Lyutikov2001,McWilliams2011,Lai2012,Piro2012,WangJS2016}.
{In particular, \cite{Chanmugam&Dulk1982} and \cite{Dulk1983} have proposed that magnetic interactions in WD-RD  systems can explain the radio emission of the magnetic cataclysmic variable, AM Her.}

This paper is organized as follows. In Section~\ref{sec:implications}, we  discuss general constraints on physical models posted by the observations of ILT J1101 + 5521 and disfavor several radiation models.
In Section~\ref{sec:unipolar}, we introduce the unipolar inductor magnetic interaction model and apply it to ILT J1101 + 5521. In Section~\ref{sec:radiation}, we propose a relativistic version of ECME as the most likely radiation mechanism and discuss possible high-energy emission within this model. In Section~\ref{sec:others} we discuss other LPRTs and suggest that the ultra-long-period sub-population of these sources may share similar physical origins. The results are summarized in Section~\ref{sec:conclusion}.

\begin{figure*}
\includegraphics[width=18 cm,height=8.5 cm]{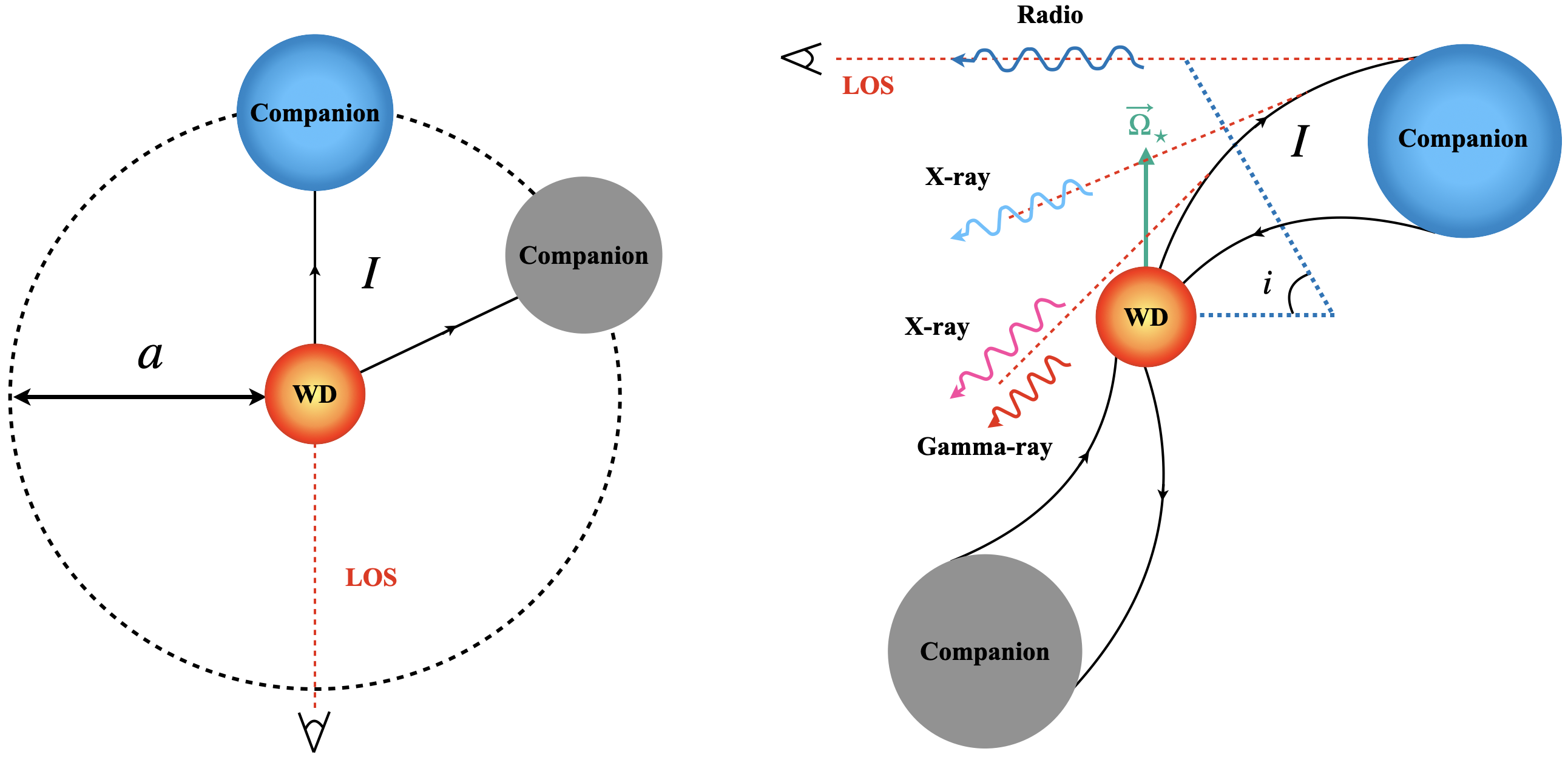}
    \caption{The geometric sketch of the WD-MD unipolar induction magnetic interaction model. The left panel is the top view and the right panel is the side view. The blue and black companion denote the configurations with and without observed radio emission, respectively. 
    The radio emission (blue wiggler) is produced by relativistic particles in the magnetic loop connecting the two stars through relativistic ECME. 
    High energy emissions are predicted, but are beamed in different directions, as marked as light blue wiggler (weak X-ray emission) via inverse Compton scattering, purple wiggler (strong X-ray emission) via relativistic boosting cyclotron radiation
    and red wiggler ($\gamma$-ray emission) via inverse Compton scattering off stellar light.
    The line of sight (LOS) is denoted as red dashed line. 
    The inclination angle $i$ is defined as the angle between the LOS of radio emission and orbital plane normal vector. The separation distance between the WD and companion star is $a$.}
    \label{fig:cartoon}
\end{figure*}

\section{Generic observational and theoretical constraints}\label{sec:implications}
We first discuss some generic constraints on any physical model based on the observational results of ILT J1101 + 5521:

(i) The observed period $\sim125.5 \ \rm min$ is consistent with the orbital period of the WD -- RD binary system, with the radio pulses observed when the binary is at inferior conjunction, i.e. the WD is between the Earth and the RD \citep{Ruiter2024}.
One can then estimate the semimajor axis of the binary as
\begin{equation}\label{eq:semimajor}
\begin{aligned}
a&=(GM)^{1/3}\left(\frac{P}{2\pi}\right)^{2/3}\\
&\simeq(5.3\times10^{10} \ {\rm cm}) \ \left(\frac{M}{0.8M_{\sun}}\right)^{1/3}\left(\frac{P}{125.5 \ \rm min}\right)^{2/3},
\end{aligned}
\end{equation}
where $M=M_\star+M_c$ is the total mass of the binary system, and the characteristic values of $M_\star=0.6M_\sun$ and $M_c=0.2M_\sun$ are taken for the WD and the RD, respectively, following \cite{Ruiter2024}. The separation between the two stars ranges from $a(1-e)$ to $a(1+e)$, where $e$ is the eccentricity. The orbit is close to a circle for this source \citep{Ruiter2024}, thus $e\simeq0$ and the distance between the two stars is $a$.

(ii) The peak flux density ranges from $\sim41 \ \rm mJy$ to $\sim256 \ \rm mJy$ with the pulse time duration $\Delta t$ lasting between 30 to 90 seconds. The observed range of the radio pulse frequency is from $120$ MHz to $168$ MHz \citep{Ruiter2024}.
The brightness temperature can be estimated based on the observed parameters as
\begin{equation}
\begin{aligned}
T_b=&\frac{S_\nu D^2}{2\pi k_B(\nu\Delta t)^2}\simeq(3.9\times10^{13} \ {\rm K}) \ \left(\frac{S_\nu}{0.1 \ \rm Jy}\right)\\
&\times\left(\frac{\nu}{140 \ \rm MHz}\right)^{-2}\left(\frac{\Delta t}{60 \ \rm s}\right)^{-2}\left(\frac{D}{504 \ \rm pc}\right)^2,
\end{aligned}
\end{equation}
where $S_\nu$ is the flux density, 
$\nu$ is the typical frequency, and $D$ is the distance.
The maximum brightness temperature for incoherent radio emission can be estimated as \citep{Zhang2023RMP}
\begin{equation}
T_{\rm b, incoh}\simeq \Gamma \gamma m_e c^2/k_B\simeq (5.9\times10^{11} \ {\rm K}) \ \Gamma\gamma_2,
\end{equation}
where $m_e$ is the electron mass, $\Gamma$ and $\gamma$ are the bulk motion Lorentz factor and the typical electron Lorentz factor, respectively. For the binary system we are interested in, there is no bulk relativistic motion of the emitter, so we only normalize the electron Lorentz factor to $100$.  Since $T_b \gg T_{\rm b,incoh}$, the radio emission must be coherent.

(iii) Since radio emission is only detected in certain orbital phases, the coherent radio emission must not be isotropically emitted. Geometric or relativistic beaming is needed (see Figure \ref{fig:cartoon} for a possible geometry). The observed emission has high polarization, with the highest linear polarization degree $\sim51\%$ for the brightest pulse and the circular polarization degree $<1.6\%$ and $<4.2\%$ for specific pulses \citep{Ruiter2024}. These polarization properties are likely produced by relativistically moving electrons.

For completeness, we list other LPRTs detected so far in Table~\ref{table}. One can see that besides ILT J1101 + 5521, a few other sources also have evidence of a possible RD companion, even though deep searches of optical counterparts of some others led to tight upper limits. X-ray emission was detected from
AR Scorpii \citep{Marsh2016,Buckley2017} and one recent source ASKAP J1832-0911 \citep{WangZT2024}.
This source was also detected by DART was named DART J1832-0911 \citep{LiD2024}.

\begin{table*}[htbp]
\centering\caption{Published LPRTs with distances, periods, flux density, luminosities, linear polarization degrees (in the radio band), possible optical counterparts and high energy (HE) counterparts. 
AR Scorpii is also included because of its possible similar origin. Corresponding references are listed below: [1]\cite{Hyman2005}, [2]\cite{Kaplan2008}, [3]\cite{Rea2022}, [4]\cite{Hurley-Walker2022}, [5]\cite{Hurley-Walker2023}, [6]\cite{Men2025}, [7]\cite{Caleb2024}, [8]\cite{Ruiter2024}, [9]\cite{Hurley-Walker2024}, [10]\cite{LiD2024}, [11]\cite{WangZT2024}, [12]\cite{Marsh2016},  [13]\cite{Buckley2017}, [14]\cite{Stanway2018}, [15]\cite{Lee2025}.}
\begin{tabular}{c|ccccccc}
\hline
\hline
Source & $D$ (kpc) & $ P \ ({\rm min})$ & $S_\nu$ (Jy) & $\Pi_L \ (\%)$ & {\rm Optical counterparts} & HE counterparts & Reference\\
\hline
GCRT J1745–3009  & $<0.07$  &77 & 1 & / & No detection & / & [1],[2] \\
\hline
GLEAM-X J1627-52 & $1.3\pm0.5$  &18.18 & 5--40 & $88\pm 1$ & No detection & / & [3],[4] \\
\hline
GPM J1839-10 & $5.7\pm2.9$ &21 &0.1--10 & 10--100 & KD or MD? & / & [5],[6] \\
\hline
ASKAPJ1935+2148 & 4.85 &53.8 &0.119 & $>90$ & near IR source ? & / & [7]\\
\hline
ILT J1101 + 5521 & 0.504 &125.5 &0.041-0.256 & $51\pm 6$ & WD+RD & / & [8]\\
\hline
GLEAM-X J 0704-37 & $1.5\pm0.5$  &174 &0.04 &20--50 & RD & / & [9]\\
\hline
DART/ASKAP J1832–0911 & 4.5  & 44 &0.5-2 & 75-100 & NS? & X-ray & [10],[11] \\
\hline
AR Scorpii & $0.116\pm0.016$  &213.6 & 0.003-0.012 & $<1\%$ & WD+RD & {X-ray} & [12],{[13],[14]}\\
\hline
ASKAP J1839-0756 & $4.0\pm1.2$ & 387 &  0.022-1.8 & 90 & No detection & / & [15] \\
\hline
\end{tabular}
\label{table}
\end{table*}

In the following, we discuss some generic constraints on theoretical models based on observational data.
The brightness temperatures of ILT J1101 + 5521 and other LPRTs are above the maximum allowed by incoherent emission, so the radio emission mechanism must be coherent. 
For ILT J1101 + 5521, we survey various emission mechanisms and can rule out the following mechanisms:

\begin{itemize}
\item 
The required specific viewing geometry and the observed high linear polarization degrees suggest that the emitting electrons should be geometrically and relativistically beamed (otherwise, radio emission should be detected in all phases). This rules out the emission mechanisms invoking non-relativistic electrons, such as the non-relativistic version of ECME.
\item 
Type III solar flares emit coherent radio emission via collective Langmuir wave plasma emission \citep{Melrose2017}. For a static plasma, the required plasma number density 
is $\sim10^7 \  \rm cm^{-3}$ in order to have $\omega_p$ reach 
the observed frequency at $\sim100$ MHz. This density is much greater than the Goldreich-Julian density of a WD magnetosphere (see Eq.~(\ref{eq:GJ-density}). 
So at least the non-relativistic plasma emission is not suitable to explain the observations. More complicated relativistic plasma emission may be possible, but it is unclear how an instability can be driven from such a system.
\item Relativistic electrons can radiate in the background magnetic field via synchrotron radiation or curvature radiation.  
For typical parameters, the emission frequencies of synchrotron radiation and curvature radiation greatly exceed the typical observed radio frequency in LPRTs.
For curvature radiation, one needs to lose the perpendicular momentum rapidly. In the characteristic emission region of interest in this paper, the background magnetic field $B_{\rm bg}$ field is not strong enough, and synchrotron cooling may not be sufficiently strong to make curvature radiation relevant. 
Also the traditional difficulties regarding formation and maintenance of the bunches also apply here. 
We therefore conclude that both synchrotron and curvature radiation are likely not the radiation mechanisms of these systems. 

\end{itemize}
In the following, we will discuss a model to interpret the LPRT radio emission. The model includes two parts. The first part (Section~\ref{sec:unipolar} discusses a well-known unipolar induction model for magnetic interaction between the two stars. The second part (Section~\ref{sec:radiation}) proposes 
a relativistic version of the ECME mechanism to account for observed coherent radio emission from ILT J1101 + 5521 and other systems with similar properties of radio emission.

\section{THE MAGNETIC INTERACTION MODEL}\label{sec:unipolar}

\subsection{Physical picture}

We consider a binary system consisting of a magnetized WD primary star with mass $M_\star$, radius $R_\star$, angular velocity $\Omega_\star$ and magnetic dipole moment $\mu=B_\star R_\star^3$, and a weakly magnetized RD with surface magnetic field strength $\leq{\rm kG}$, with mass $M_c$ and radius $R_c$. The distance between the two stars is $a$. The angular velocity of the WD is $\Omega_\star = \xi \Omega$, where the orbital angular velocity is $\Omega = 2\pi/P$ 
($P$ is the orbital period) {and $\xi \equiv \Omega_\star/\Omega$ is a factor describing the ratio between the two angular velocities}.

We first show that the WD carries a force-free magnetosphere. 
The gravitational force on a proton at the WD surface can be estimated as
\begin{equation}
\begin{aligned}
F_{\rm grav}&=\frac{GM_\star m_p}{R_\star^2}\\
&\simeq(2.7\times10^{-16} \ {\rm dynes}) \ \left(\frac{M_\star}{0.6M_\sun}\right)\left(\frac{R_\star}{0.01R_\sun}\right)^{-2},
\end{aligned}
\end{equation}
where $m_p$ is the proton mass.
The electric force, on the other hand, may be calculated as
\begin{equation}
\begin{aligned}
F_{\rm EM}&=\frac{q\Omega_\star R_\star B_\star}{c}\\
&\simeq(9.3\times10^{-9} \ {\rm dynes}) \ \xi \left(\frac{P}{125.5 \ \rm min}\right)^{-1}\left(\frac{R_\star}{0.01R_\sun}\right)\left(\frac{B_\star}{10^6 \ \rm G}\right).
\end{aligned}
\end{equation}
One can see that $F_{\rm EM}\gg F_{\rm grav}$ is satisfied, justifying a magnetically dominated magnetosphere similar to pulsar magnetospheres. 
The radius of the WD magnetosphere light cylinder can be estimated as 
\begin{equation} 
R_{\rm LC}=\frac{c}{\Omega_\star}\simeq(3.6\times10^{13} \ {\rm cm}) \ \xi^{-1}\left(\frac{P}{125.5 \ \rm min}\right) \gg a
\end{equation}
if $\xi \ll 10^3$. This suggests that the companion MD star is most likely inside the magnetosphere of the WD. 
The surface magnetic field of a typical RD is several kG from radio observations of magnetic cataclysmic variables (MCVs) \citep{Barrett2020}. 
For an RD with radius $R_c=0.2R_{\sun}$,
the magnetic field strengths from the two stars are equal to $\sim18 \ {\rm G}$ at $r_c \simeq2.67\times10^{10} \ \rm cm$ away from the WD, 
which is close to the separation distance ($a\simeq5.3\times10^{10} \rm cm$). 
As long as the radio emission region is at $r<r_c$ from the WD (which is the case for the geometric configurations we consider), 
the WD magnetic field is the dominant component in the magnetosphere.

We consider a toy model with $\bf \Omega_\star$ and $\bf \Omega$ aligned. The MD moving with respect to the WD magnetosphere would generate an electric potential drop of the magnitude of
$\Phi\simeq 2R_c|\vec E|$ due to the unipolar induction effect. Here the factor of two is included since the circuit is repeated in both hemispheres. The induced electric field is
\begin{equation}
\vec E=\frac{\vec v}{c}\times\vec B_{\rm WD},
\end{equation}
where $\vec v = |\vec\Omega_\star - \vec\Omega| \times  \vec r = \Delta\Omega  r \hat{\phi}$ is the companion star’s velocity relative to the WD magnetosphere, $\hat{\phi}$ is the unit vector along azimuthal direction, $\vec B_{\rm WD}$ is the background magnetic field of the WD and
\begin{equation}
\Delta\Omega=|\Omega-\Omega_\star|=|\Omega-\xi\Omega|=|1-\xi|\Omega=\zeta\Omega.
\end{equation}
We can then obtain the total voltage when the two stars are at a separation of $a$  \citep{Lai2012,Piro2012}
\begin{equation}
\begin{aligned}
\Phi\simeq \frac{2\mu R_c}{ca^2}\Delta\Omega\simeq&(9.1\times10^7 \ {\rm statvolt}) \ \zeta\left(\frac{B_\star}{10^6 \ \rm G}\right)\left(\frac{R_c}{0.2R_\sun}\right)\\
&\times\left(\frac{M}{0.8M_{\sun}}\right)^{-2/3}\left(\frac{P}{125.5 \ \rm min}\right)^{-7/3},
\end{aligned}
\end{equation}
{The maximum Lorentz factor achievable by an electron accelerated in this potential can be estimated as}
\begin{equation}
\begin{aligned}
\gamma_{\rm max}=\frac{q\Phi}{m_ec^2}&\simeq5.4\times10^4 \ \zeta\left(\frac{B_\star}{10^6 \ \rm G}\right)\left(\frac{R_c}{0.2R_\sun}\right)\\
&\times\left(\frac{M}{0.8M_{\sun}}\right)^{-2/3}\left(\frac{P}{125.5 \ \rm min}\right)^{-7/3}.
\end{aligned}
\label{eq:gamma_max}
\end{equation}
Notice that the star can spin faster than the orbit.   
{Unless under the synchronous condition due to tidal lock \citep{Wu&Wickramasinghe1993}, $\zeta>0$ is usually satisfied because $\Delta\Omega$ is defined as the absolute value of $\Omega_\star$ and $\Omega$ difference. }
The current in the circuit may be calculated as
$I={\Phi}/{2\mathcal{R}_{\rm mag}}$, where
$\mathcal{R}_{\rm mag}={4\pi}/{c}$
is the resistance of the magnetosphere. One can then estimate the total electric power dissipation rate of the binary system as \citep{Lai2012,Piro2012}
\begin{equation}
\begin{aligned}
\Dot{E}&_{\rm diss}=\frac{2\Phi^2}{\mathcal{R}_{\rm mag}}
\simeq(2.0\times10^{25} \ {\rm erg \ s^{-1}}) \ \zeta^2\left(\frac{B_\star}{10^6 \ {\rm G}}\right)^2\\
&\times\left(\frac{R_c}{0.2R_\sun}\right)^2\left(\frac{M}{0.8M_{\sun}}\right)^{-4/3}\left(\frac{P}{125.5 \ \rm mins}\right)^{-14/3}.
\end{aligned}
\end{equation}
The highest observed specific radio luminosity is $\sim7.8\times10^{19} \ \rm erg \ s^{-1} \ Hz^{-1}$ \citep{Ruiter2024}. 
One can see that the model can satisfy this constraint if $\zeta>10$.

\cite{Chanmugam&Dulk1982} and \cite{Dulk1983} assumed a near synchronous configuration ($\zeta \ll 1$) in their model to interpret radio emission of the magnetic cataclysmic variable, AM Her. Their typical Lorentz factor is $\sim 2$ because of a much smaller electric potential. In our model, we assume that the two stars are far from tidally locked and allow $\zeta$ as a free parameter. As long as $\zeta \gg 0$, electrons can be highly relativistic. It is worth mentioning that magnetized cataclysmic variables (mCVs), which are known sources of WD-RD systems with accretion onto the surface of WDs, typically have either $\zeta \gg 1$ (for the so-called intermediate-polars) when accretion is significant, or $\zeta \sim 0$ (for the so-called polars), when accretion does not exist and the system undergoes tidal locking \citep{warner95}. The coherent radio emission from LPRTs suggests that these sources should not have accretion (otherwise coherent emission would be quenched, as is the case of radio pulsars, \citealt{papitto13}). Indeed, these sources do not show observational signatures of known mCVs. They may be in the short transition phase from intermediate polars to synchronized systems, allowing the $\zeta$ value varying in a wide range between 0 and 10. In the rest of the discussion, we normalize $\zeta$ to unity but keep in mind that it could be an arbitrary value in a wide range. The unipolar mechanism we propose applies to a wide range of $\zeta$ values.

\subsection{Geometric configuration}

Within this model, the particle flow that powers radio emission should be confined in the magnetic loop connecting the two stars. As a result, coherent radio emission can only be detected when the magnetic loop plane aligns with the line of sight, i.e. at conjunction (see Figure~\ref{fig:cartoon} left panel for the top view of the orbit). In order to suppress emission from other phases, the radiating electrons are required to move relativistically, consistent with the estimate in Eq.~(\ref{eq:gamma_max}). 
{The observation of radio emission when the MD is at superior conjunction suggests that electrons should move towards the WD direction with the parallel electric field direction pointing away from the WD. Such a configuration can be only achieved if the WD is an anti-parallel rotator, i.e. ${\bf \Omega \cdot \mu} < 0$ \citep{Ruderman1975}.}

The conjunction configuration alone cannot guarantee the detection of radio emission. Because of the relativistic motion of the particles, the emission beam is essentially along the tangential direction of the field lines. 
Radio emission can be observed only when the magnetic field line tangential directions are close to the observer.
Figure~\ref{fig:cartoon} right shows the side view of the system, with a marginal case that the field tangential direction barely reaches the line of sight. When the orbital inclination angle $i$ is even smaller (e.g. close to the face on configuration), no coherent radio emission can reach the observer. For ITL J1101+5521, optical observations give the constraint of $i < 74^{\circ}$ and $i>50^\circ$ from the observed radial velocity amplitude \citep{Ruiter2024}. 
Our model provides another limit from the low end, with the minimum $i$ depending on the unknown angle between $\bf \Omega_\star$ and $\bf \mu$.

\section{Radiation Mechanism}\label{sec:radiation}

In this section, we investigate the relativistic ECME mechanism for coherent radio emission and the possible associated high energy counterparts including X-ray and $\gamma$-ray emission through relativistic boosting cyclotron radiation and inverse Compton scattering, respectively.

\subsection{Characteristic frequencies} 

We first study the relevant physical parameters in the WD magnetosphere.
Since the WD magnetosphere can be treated as force-free, we can define a
Goldreich–Julian (GJ) density \citep{Goldreich&Julian1969} for the WD
\begin{equation}\label{eq:GJ-density}
\begin{aligned}
n_{\rm GJ}=&\frac{B_\star\Omega_\star}{2\pi qc}\left(\frac{r}{R_\star}\right)^{-3}\simeq(1.6\times10^{-4} \ {\rm cm^{-3}}) \ \left(\frac{B_\star}{10^6 \ \rm G}\right)\\
&\times\left(\frac{P}{125.5 \ \rm min}\right)^{-1}\left(\frac{r}{2.67\times10^{10} \ \rm cm}\right)^{-3}\left(\frac{R_\star}{0.01R_\sun}\right)^3.
\end{aligned}
\end{equation}
The total number density may be written as $n=\kappa n_{\rm GJ}$, where $\kappa$ is the multiplicity parameter.
The plasma frequency corresponding to the GJ-density can be estimated as
\begin{equation}
\begin{aligned}
&\omega_p=\sqrt{\frac{4\pi q^2n}{\gamma m_e}}\simeq(2.3\times10^2 \ {\rm rad \ s^{-1}}) \ \kappa^{1/2}\left(\frac{B_\star}{10^6 \ \rm G}\right)^{1/2}\\
&\times\left(\frac{P}{125.5 \ \rm min}\right)^{-1/2}\left(\frac{r}{2.67\times10^{10} \ \rm cm}\right)^{-3/2}\left(\frac{R_\star}{0.01R_\sun}\right)^{3/2}\left(\frac{\gamma}{10}\right)^{-1/2}.
\end{aligned}
\end{equation}
On the other hand, the electron cyclotron frequency can be estimated as
\begin{equation}
\begin{aligned}
\omega_B=\frac{eB_\star}{m_ec}\left(\frac{R_\star}{r}\right)^{3}\simeq&(3.1\times10^{8} \ {\rm rad \ s^{-1}}) \ \left(\frac{B_\star}{10^6 \ \rm G}\right)\\
&\times\left(\frac{R_\star}{0.01 R_\sun}\right)^{3}\left(\frac{r}{2.67\times10^{10} \ \rm cm}\right)^{-3}.
\end{aligned}\label{Eq:omegaB}
\end{equation}
For typical parameters, one has $\omega>\omega_B \gg\omega_p$.

\subsection{Relativistic electron cyclotron maser emission}

We believe that a relativistic version of the ECME mechanism is likely operating in the system to power the observed coherent radio emission. 
An anisotropic momentum distribution of relativistic plasma might be generated due to synchrotron radiation cooling \citep{Bilbao2024}.
The ECME mechanism \citep{Twiss1958,Wu&Lee1979} has been widely accepted to interpret decametric emission from Jupiter, auroral kilometric radiation from Earth, and magnetic cataclysmic variables. The Jupiter emission is triggered by the similar unipolar induction mechanism due to Jupiter-Io interaction \citep{Goldreich&Lynden-Bell1969}.
Three physical conditions are required to produce ECME: The first is population inversion of the energetic electrons in the direction perpendicular to the background magnetic field to achieve negative absorption, i.e. $\partial f/\partial p_\perp>0$, where $f$ is the particle distribution function and $p_\perp$ is the momentum perpendicular to the background magnetic field. 
The second condition requires $\omega_B\gg \omega_p$, which is satisfied here. Finally, one should satisfy the gyromagnetic resonance condition, which can be written as \citep{Wu&Lee1979,Melrose2017}
\begin{equation}
\begin{aligned}
\omega&=\frac{s\omega_B}{\gamma\left(1-\frac{v_\parallel}{c}\cos\theta_v\right)}\\
&=s\frac{\omega_B}{\gamma}+k_\parallel v_\parallel, \ s=0,\pm 1,\pm 2 \dots,
\end{aligned}
\end{equation}
where $s$ is the harmonic number, $k_\parallel$ and $v_\parallel$ denote the wave vector and velocity of the electron along the background magnetic field. 
We consider the case of $s=1$. 
Thus the emission frequency of relativistic version of ECME can be estimated as
\begin{equation}
\begin{aligned}
\nu&=\delta \frac{\omega_B}{2\pi}
\simeq\frac{\gamma\omega_B}{2\pi}\simeq(5.0\times10^8 \ {\rm Hz})\\
&\times\left(\frac{r}{2.67\times10^{10} \ \rm cm}\right)^{-3}\left(\frac{R_\star}{0.01R_\sun}\right)^{3}\left(\frac{B_\star}{10^6 \ \rm G}\right),
\end{aligned}
\end{equation}
where $\delta=1/[\gamma(1-\beta\cos\theta_v)]$ is the Doppler factor, which $\simeq \gamma$ when $\theta_v \sim 1/\gamma$, $\beta=v_\parallel/c$ is the normalized velocity and its direction is along the background magnetic field.
It should be pointed out that the magnetic field strength, Lorentz factor of the electron and viewing angle can significantly influence the emission frequency. 
One can see that for typical parameters, the emission frequency is consistent with the typical frequency of several hundred MHz as observed in LPRTs.

\begin{figure}
\includegraphics[width=\columnwidth]{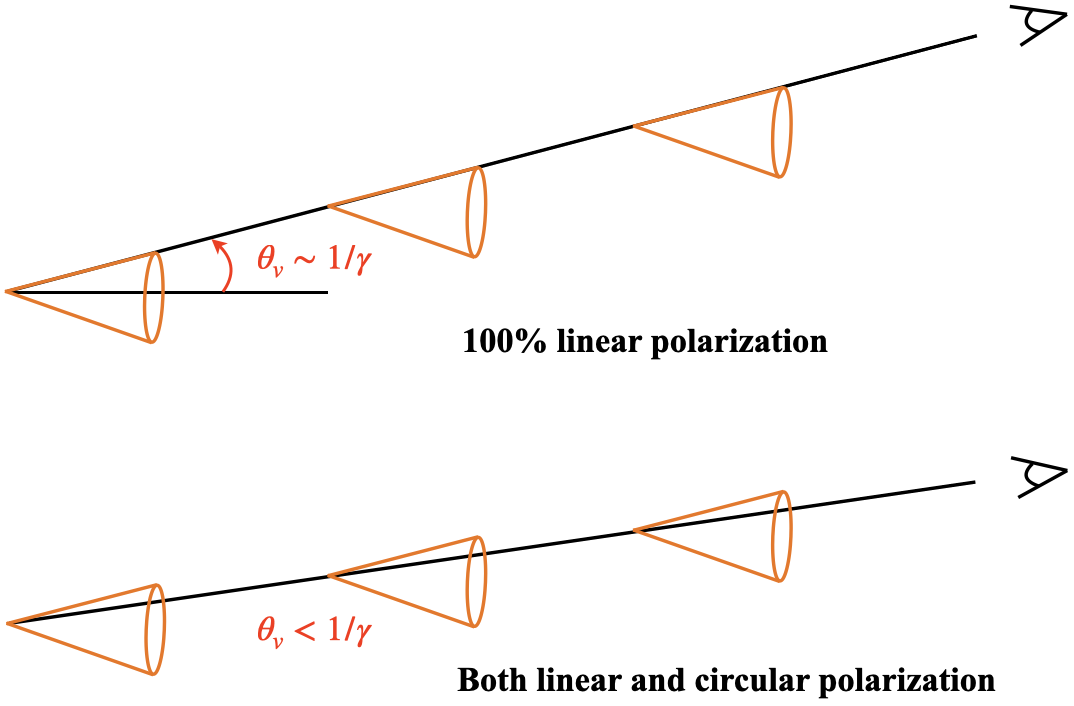}
    \caption{A cartoon picture for the relativistic ECME. The solid line denotes the line of sight. The main radiation cone is denoted as the orange cone with half opening angle $\sim 1/\gamma$. }
    \label{fig:polarization}        
\end{figure}

The observed emission of ILT J1101+5521 is mostly linearly polarized but also carries circular polarization. {As shown in Figure~\ref{fig:polarization}, strong linear polarization with partial circular polarization is naturally expected for relativistic ECME. In Appendix \ref{app:polarization}, we present a calculation of cyclotron radiation of one single particle in the comoving frame. We show that 100\% linear polarization can be observed only in the direction perpendicular to the magnetic field line and circular polarization will show up in other viewing angles (see Eq.~(\ref{eq:polarization degree})). For relativistic ECME, the brightest emission comes from the $1/\gamma$ cone, which corresponds to the $90^{\rm o}$ viewing angle in the comoving frame. After Lorentz transformation, 100\% linear polarization can be observed when the line of sight is along the $1/\gamma$ viewing angle at the edge of the radiation cone in the lab frame (upper panel of Figure~\ref{fig:polarization}). More generally, when the line of sight is within the $1/\gamma$ cone (lower panel of Figure~\ref{fig:polarization}), mixed linear and circular polarization can be observed.}

We invoke a relativistic version of the ECME mechanism with electrons moving with $\gamma \sim 10$ to account for the emission of ILT J1101+5521. This Lorentz factor is smaller than the maximum Lorentz factor achievable due to unipolar induction (Eq.~(\ref{eq:gamma_max})) so is reasonable. 
The direction of the dominant linear polarization is at $1/\gamma \sim 0.1$ with respect to the electron motion direction. The observation is consistent with the line of sight slightly deviates from the tangential direction of the magnetic field lines.
The non-negligible circular polarization suggests that the line of sight is slightly inside the $1/\gamma$ cone.

The observed bandwidth of the brightest pulse in ILT J1101 + 5521 is $\sim40 \ \rm MHz$ which is extremely narrow but still not monochromatic. Since cyclotron emission has a characteristic frequency, one expects that the ECME mechanism carries a narrow spectrum. Relativistic ECME introduces a spectral broadening if the $\gamma$ distribution has a spread around a central value. The radiation flux drops significantly both at much higher or lower frequencies. This is consistent with observations.

\subsection{High energy counterparts}\label{sec:high-energy}

Since relativistic particles are involved, this model also predicts high-energy emissions.
In the following, we discuss possible high-energy radiation mechanisms and estimate the X-ray  emission power. 
When electrons are accelerated along the background magnetic field and move towards the WD, 
both electron Lorentz factor and background magnetic field strength of the WD increase.
However, since electrons are moving nearly along the field line with an extremely small pitch angle, synchrotron radiation cannot occur. 
In the rest frame without parallel motion, the electron is still non-relativistic and radiates cyclotron emission.
The maximum observed radiation frequency is boosted by the maximum Lorentz factor (Eq.~(\ref{eq:gamma_max})), i.e. 
\begin{equation}
\begin{aligned}
&\omega_{\rm cyc}\simeq\gamma_{\rm max}\omega_B\simeq (3.2\times10^{17} \ {\rm rad \ s^{-1}}) \ \zeta\left(\frac{B_\star}{10^6 \ \rm G}\right)^2\left(\frac{R_c}{0.2R_\sun}\right)\\
&\times\left(\frac{M}{0.8M_{\sun}}\right)^{-2/3}\left(\frac{P}{125.5 \ \rm min}\right)^{-7/3}\left(\frac{r}{10^9 \ \rm cm}\right)^{-3}\left(\frac{R_\star}{0.01R_\sun}\right)^{3},
\end{aligned}
\end{equation}
which is in the soft X-ray band. The X-ray emission is beamed, and usually deviate from the direction of radio emission (Figure~\ref{fig:cartoon}).

The total number of observable emitting charged particles at any instant can be estimated as $N_{\rm tot}\sim \pi (l_e/\gamma)^2 l_e n$, where $l_e$ is the typical length scale of the radiation region,
noticing that the cross section of the emission beam is defined by $l_e/\gamma$ due to relativistic beaming. 
The received power is boosted by a factor of $\gamma^2$ since the observed time interval is shorter by a factor of $1-\beta\cos\theta_v\sim 1/\gamma^2$ with respect to the emission time interval.
Consider that the electron moves non-relativistically in the perpendicular direction with normalized velocity $\beta\sim0.5$, the maximum X-ray luminosity due to relativistically boosted cyclotron radiation can be calculated as
\begin{equation}
\begin{aligned}
L_{\rm X,cyc}^{\rm max}&\simeq \gamma_{\rm max}^2 N_{\rm tot} r_0^2c\beta^2B_{\rm bg}^2\\
&\simeq (1.3\times10^{27} \ {\rm erg \ s^{-1}}) \ \kappa\zeta\left(\frac{B_\star}{10^6 \ \rm G}\right)^5\left(\frac{R_c}{0.2R_\sun}\right)\\
&\times\left(\frac{M}{0.8M_{\sun}}\right)^{-2/3}\left(\frac{P}{125.5 \ \rm min}\right)^{-10/3}\left(\frac{l_e}{10^{10} \ \rm cm}\right)^3\\
&\times\left(\frac{R_\star}{0.01R_\sun}\right)^9\left(\frac{r}{10^9 \ \rm cm}\right)^{-3}\left(\frac{\beta}{0.5}\right)^2,
\end{aligned}
\label{eq:Lcyc}
\end{equation}
where $r_0=e^2/m_ec^2$ is the classical electron radius.

Two LPRTs have X-rays detected. AR Scorpii with a 213.6-min period has an X-ray luminosity of $\sim5\times10^{30} \ {\rm erg \ s^{-1}}$. Such a luminosity is reachable if AR Scorpii WD has a stronger magnetic field (note the steep dependence $P_{\rm cyc}\propto  B_\star^5$ in Eq.~(\ref{eq:Lcyc})). Recently, another LPRT DART/ASKAP J1832–0911 with a $44.2$-min period was detected in X-rays with a high luminosity of $\sim10^{33} \ {\rm erg \ s^{-1}}$ \citep{WangZT2024}.  
Such a high luminosity is difficult to reach within our model unless $B_*$ is extremely high. This source on the other hand has a supernova remnant association \citep{LiD2024}, suggesting a different (e.g. young magnetar) origin.

X-rays and even $\gamma$-rays can also be produced via inverse Compton (IC) scattering of relativistic particles off the thermal photons from the stars. The surface temperature of a WD is $T=(10^4 \ {\rm K}) \ T_4$. At a distance $r$ from the WD, the photon energy density is  $U_{\rm ph}=U_{\rm ph,\star}(r/R_\star)^{-2} = a_{\rm rad}T^4 (r/R_\star)^{-2}$.  
The surface temperature of the WD with $T=10^4 \ \rm K$ defines a peak photon energy $h\nu=2.82k_BT$ with corresponding peak frequency $\nu\simeq 5.9\times10^{14} \ {\rm Hz} \ T_4$. The typical frequency of the IC photons can be estimated as
\begin{equation}
\nu_{\rm ICS}\simeq \gamma^2\nu\simeq(5.9\times10^{18} \ {\rm Hz}) \ \left(\frac{\gamma}{10^2}\right)^2\left(\frac{T}{10^{4} \ \rm K}\right),
\end{equation}
which is in the X-ray band. 
Note that we have adopted a higher Lorentz factor $\gamma \sim 10^2$ than the one to account for radio emission ($\gamma\sim 10$). 
This is possible because electrons are continuously accelerated by the parallel electric field along the magnetic field line.
Since magnetic field lines are curved, the X-ray emission, emitted at a different location, would be beamed in a somewhat different direction (see Figure~\ref{fig:cartoon}) and can be detectable only under preferred viewing geometry. 

The total observed X-ray IC luminosity can be estimated as
\begin{equation}
\begin{aligned}
L_{\rm X,IC}&\simeq \gamma^2 N_{\rm tot}\gamma^2\sigma_{\rm T}cU_{\rm ph}\\
&\simeq(7.1\times10^{22} \ {\rm erg \ s^{-1}}) \ \kappa\left(\frac{\gamma}{10^2}\right)^{2}\left(\frac{r}{10^9 \ \rm cm}\right)^{-5}\left(\frac{T}{10^4 \ \rm K}\right)^4\\
&\times\left(\frac{l_e}{10^{10} \ \rm cm}\right)^{3}\left(\frac{R_\star}{0.01R_\sun}\right)^{5}\left(\frac{B_\star}{10^6 \ \rm G}\right)\left(\frac{P}{125.5 \ \rm min}\right)^{-1},
\end{aligned}
\end{equation}
where $\sigma_{\rm T}$ is the Thomson cross section and one can see that the observed luminosity is much smaller than that due to relativistically boosted cyclotron radiation (Eq.~(\ref{eq:Lcyc})).

As the electrons further accelerated toward the WD surface, Lorentz factor further increases. 
For the most extreme case, we assume that the electron Lorentz factor reaches the maximum ($\gamma\sim10^5$) near the WD. 
The IC photon frequency becomes
\begin{equation}
\begin{aligned}
\nu_{\rm ICS}&\simeq\gamma^2\nu\simeq(5.9\times10^{24} \ {\rm Hz}) \ \left(\frac{\gamma}{10^5}\right)^2\left(\frac{T}{10^{4} \ \rm K}\right),
\end{aligned}
\end{equation}
which is in the $\gamma$-ray band. 
The corresponding observed $\gamma$-ray luminosity is 
\begin{equation}
\begin{aligned}
L_{\rm \gamma,IC}&\simeq \gamma^2N_{\rm tot}\gamma^2\sigma_{\rm T}cU_{\rm ph}\\
&\simeq(7.1\times10^{28} \ {\rm erg \ s^{-1}}) \ \kappa\left(\frac{\gamma}{10^5}\right)^{2}\left(\frac{r}{10^9 \ \rm cm}\right)^{-5}\left(\frac{T}{10^4 \ \rm K}\right)^4\\
&\times\left(\frac{l_e}{10^{10} \ \rm cm}\right)^{3}\left(\frac{R_\star}{0.01R_\sun}\right)^{5}\left(\frac{B_\star}{10^6 \ \rm G}\right)\left(\frac{P}{125.5 \ \rm min}\right)^{-1}.
\end{aligned}
\end{equation}
The $\gamma$-ray flux is typically too low to be detected. Under special geometric configurations, $\gamma$-rays may be beamed to the WD and be reprocessed as thermal emission. Assuming that the thermal emission is re-emitted in a hot spot with radius $l \sim 0.01 R_*$, the temperature is 
\begin{equation}
\begin{aligned}
T_{\rm max,spot}&\simeq\left(\frac{L_{\rm \gamma,IC}}{\pi l^2\sigma}\right)^{1/4}\simeq(5.3\times10^4 \ {\rm K}) \ \\
&\times\left(\frac{P_{\rm ICS}}{7\times10^{28} \ {\rm erg \ s^{-1}}}\right)^{1/4}\left(\frac{l}{0.01R_\star}\right)^{-1/2},
\end{aligned}
\end{equation}
which is of the same order of the surface temperature of the WD. Here $\sigma$ is the Stefan–Boltzmann constant. If the heat is instead spread out to the full WD surface, the corresponding temperature is
\begin{equation}
\begin{aligned}
T_{\rm max,full}&\simeq\left(\frac{L_{\rm \gamma,IC}}{4\pi R_\star^2\sigma}\right)^{1/4}\simeq(3.8\times10^3 \ {\rm K}) \ \\
&\times\left(\frac{P_{\rm ICS}}{7\times10^{28} \ {\rm erg \ s^{-1}}}\right)^{1/4}\left(\frac{l}{0.01R_\star}\right)^{-1/2},
\end{aligned}
\end{equation}
which is below the WD surface temperature. In any case, the effect of this heating effect is likely undetectable.

\section{Other sources}\label{sec:others}

We have shown that the proposed magnetic interaction model invoking a WD -- MD binary can successfully explain the observations of ILT J1101+5521. We suspect that the same model may apply to at least some of the other LPRTs as well, but probably with different parameters and observational geometric configurations. 
In the following, we discuss whether other reported LPRTs and related objects (summarized in Table~\ref{table}) align with the framework of our model:

\begin{itemize}
\item Besides ILT J1101 + 5521 \citep{Ruiter2024}, the most likely candidate that has the similar origin is the recently detected GLEAM-X J 0704-37 with a 2.9-hour period \citep{Hurley-Walker2024}. The polarization behavior of the source is variable with non-negligible linear (20–50\%) and circular polarization (10–30\%) over millisecond timescales. Optical observations revealed an MD counterpart \citep{Hurley-Walker2024}. Deeper observations will reveal whether the MD is in a binary system with a WD companion. 
The observed mixed linear and circular polarization can be produced when the line of sight is inside the $1/\gamma$ cone in this model.
\item A very relevant system is the optical white dwarf pulsar AR Scorpii with a period of 213.6 minutes \citep{Marsh2016,Buckley2017}.
The observed broadband spectra of AR Scorpii at high frequency band is consistent with synchrotron radiation via relativistic electrons. 
The polarization properties of AR Scorpii is quite different from other intermediate polars with dominant circular polarization which might be produced through cyclotron emission.
The optical observations of AR Scorpii show that the linear polarization ($\Pi_L\sim40\%$) and both low circular polarization ($\Pi_V$ is a few percent) are consistent with synchrotron emission \citep{Buckley2017}, probably originates from a bow shock at the interface of the WD and MD winds \citep{Geng2016}.
Radio emission from AR Scorpii seems to have two origins. Above 10 GHz, the radio emission is not polarized and it is consistent with self-absorbed synchrotron radiation. Below 10 GHz, it shows weak linear polarization and a circular polarization with $\Pi_V\sim30\%$. 
Its pulsation follows the beat frequency instead of the orbital frequency, which is different from ILT J1101+5521. It was suggested that this emission originates from non-relativistic ECME \citep{Stanway2018}. Within our framework, the beamed relativistic ECME emission in the current loop region might be missed due to a unfavorable geometry, and the observed radio emission may be related to reprocessed emission from the a non-relativistic plasma outside of the current loop region.

\item The first long-period radio source was a transient source named  GCRT J1745–3009 with a period of 77 minutes \citep{Hyman2005}. Motivated by the suggestion that the source could be a WD pulsar in the pair-production death valley \citep{ZhangGil2005}, \cite{Kaplan2008} performed a deep search in the optical/infrared band and placed a 3$\sigma$ limiting magnitudes from four bands: $I>26$, $J>21$, $H>20$ and $K>19$. This rules out the existence of a WD if the source distance is $<1$ kpc. This source can be understood within our model if it is located at a distance larger than 1 kpc. The source was only briefly detected for a few periods. It could be understood if the line of sight is barely grazing the magnetic loop. Re-detection of the source during its next reactivation phase will help to test this model. 

\item GLEAM-X J1627-52 with a period of 18.18 minutes \citep{Hurley-Walker2022} and GPM J1839-10 with a period of 21 minutes \citep{Hurley-Walker2023,Men2025} are two relatively fast LPRTs. Their emissions have polarization properties more similar to pulsars, e.g. high linear polarization ($\Pi_L \sim88\%$) for GLEAM-X J1627-52 \citep{Hurley-Walker2022} and $\Pi_L \sim 10\%-100\%$ for GPM J1839-10 \citep{Hurley-Walker2023}, as well as flat polarization angle with occasional $90^\circ$ jumps. Deep searches for optical counterparts gave inconclusive results \citep{Rea2022,Hurley-Walker2023}. These two sources are least likely consistent with our WD -- RD interaction model and may have a different origin.
However, if these sources are interpreted within our model, the observed high linear polarization requires that the line of sight is close to the edge of the radiation cone ($\theta_v\sim 1/\gamma$).

\item ASKAPJ193505.1+214841.0 was reported to have a period of 53.8 minutes \citep{Caleb2024}. The polarization properties are highly variable: a bright pulse state with highly linear polarization ($>90\%$) and a weak pulse state ($\sim$26 times fainter than the bright state) with highly circular polarization. 
Within this model, the two emission components correspond to two emission regions with different viewing angles, with the bright component observed near the radiation cone edge and the fainter component observed within the radiation cone.
There is a near IR source in the field, but whether it is associated with the radio emitter is uncertain. Further observations are needed to establish whether the IR source is an RD associated with the LPRT, and whether the source could be understood within the framework of the WD -- RD interaction model.

\item 
DART J1832-0911 and ASKAP J1832-0911 are the same LPRT source with a 44.2-minute period detected by DAocheng Radio Telescope (DART) and Australian SKA Pathfinder (ASKAP), respectively \citep{LiD2024,WangZT2024}.
The source was found to be associated with a supernova remnant, favoring a young neutron star origin \citep{LiD2024}.
The radio emission is $\sim 100\%$ linear polarized with a relatively flat polarization angle across the pulses and the frequency bandwidth is relatively broad. \citep{LiD2024}.
The radio emission polarization properties are also reported to be highly linear polarized with a total polarization degree $\Pi_p\simeq92\pm3\%$. 
Linear polarization is $\Pi_L\sim 75\%$ and circular polarization is $\Pi_V\sim 50\%$ \citep{WangZT2024}.
The source also has an X-ray emission counterpart with luminosity $\sim10^{33} \ {\rm erg \ s^{-1}}$ \citep{WangZT2024}.
This source is very likely related to a newborn neutron star, not related to the mechanism discussed in this paper.

\item 
ASKAP J183950.5-075635.0 was reported to have a period of 387 minutes \citep{Lee2025}. 
The emission is highly polarized with a linear polarization degree $\sim90\%$ and a circular polarization degree $\sim37\%$. This source can be interpreted within the framework of our model, with the line of sight slightly within the $1/\gamma$ cone. The narrow spectrum is also consistent with the prediction of the relativistic ECME model.

\end{itemize}

\section{Conclusions}\label{sec:conclusion}

In this paper, we propose a magnetic interaction model invoking a magnetized WD and a weakly magnetized RD through the unipolar induction effect to interpret the newly discovered LPRT, ILT J1101+5521. We argue that a relativistic version of the ECME is the most likely emission mechanism. This is a more energetic version of the Jupiter-Io interaction observed in the solar system.

We further explore the possibility of applying this model to the emergent LPRT population with a rapidly growing number. 
We find that those with relative long periods such as GLEAM-X J0704-37 and GCRT J1745-3009 and probably ASKAPJ193505.1+214841.0 and ASKAP J183950.5-075635.0 as well, may be interpreted within the same theoretical framework. 
The WD optical pulsar AR Scorpii may also belong to the same type of system with a different viewing geometry.
The asynchronous configuration indicates large electric potential and relativistic particles. Relativistic ECME can produce both high linear polarization (if the line of sight is close to the $1/\gamma$ cone) and high circular polarization (if the line of sight is well within the $1/\gamma$ cone).

On the other hand, we note that the sources with relatively shorter periods, such as GLEAM-X J1627-52, GPM J1839-10, and DART/ASKAP J1832-0911, may have a different physical origin, e.g. a single WD pulsar or a long-period magnetar. More multi-wavelength observations of these systems and new sources will help to test the suggested dichotomy and the unified picture of the ultra-long period sub-population of LPRTs. The supernova remnant association with DART/ASKAP J1832-0911 \citep{LiD2024} provides the first observational clue of such dichotomy.

We also predict high-energy emissions from LPRTs within the framework of our model. X-rays can be produced via relativistically boosted cyclotron radiation with the maximum Lorentz factor and both X-rays and $\gamma$-rays can be produced via IC scattering of relativistic particles off the thermal photons from WD surface. 
We find that the observed X-ray counterpart of AR Scorpii can be achieved in the WD-RD system, but that of and DART/ASKAP J1832-0911 may originate from other systems, possibly invoking a newborn magnetar. In the future, the detection of X-ray counterparts and possibly $\gamma$-ray counterparts could be used to distinguish different engines of LPRTs.

\section*{Acknowledgements}
We thank Manisha Caleb, Di Li, Yunpeng Men, Nanda Rea, Kaustubh Rajwade, and Ziteng Wang for discussion and an anonymous referee for very helpful comments.
This work is supported by the Nevada Center for Astrophysics and NASA 80NSSC23M0104.

\appendix

{

\section{Polarization of relativistic ECME}\label{app:polarization}

In this appendix, we provide a calculation of the polarization properties of relativistic ECME. 
We first treat the problem in the comoving frame of the relativistic electron that moves along background magnetic field with Lorentz factor $\gamma$, and then perform Lorentz transformation to study radiation back to the lab frame.
In the comoving frame, the electron gyrates around the background magnetic field and experiences cyclotron or synchrotron radiation.

Consider an electron gyrates around the background magnetic field non-relativistically and emits cyclotron radiation. We define the viewing angle $\theta_v'$ between the line of sight and background magnetic field. 
The circular and linear polarization degrees can be written as
\begin{equation}\label{eq:polarization degree}
\Pi_{V,\rm cyc}=-\frac{2\cos\theta_v'}{1+\cos^2\theta_v'} \  \ \& \ \  \Pi_{L,\rm cyc}=\frac{1-\cos^2\theta_v'}{1+\cos^2\theta_v'}.
\end{equation}
One can see that when the line of sight is parallel to the gyration plane, i.e. $\theta_v'=\pi/2$, the radiation is 100\% linearly polarized in the comoving frame. 
When the line of sight is along the background magnetic field, i.e. $\theta_v'=0$, the radiation is 100\% circularly polarized.

Now consider emission in the lab frame. The Lorentz transformation on the viewing angle is 
\begin{equation}
\sin\theta_v'={\cal D}\sin\theta_v, \ {\cal D}=\frac{1}{\gamma(1-\beta\cos\theta_v)},
\end{equation}
where ${\cal D}$ is the Doppler factor, $\gamma$ is the Lorentz factor of the electron moving along the background magnetic field.
When $\gamma\gg1$, $\theta_v'=\pi/2$ corresponds to $\theta_v\sim 1/\gamma$. Therefore, when the line of sight is along $\theta_v\sim1/\gamma$, the observer will detect $\sim100\%$ linear polarization. 
When $\theta_v=0$, the observer will detect $\sim100\%$ circular polarization. More generally, the radiation is elliptically polarized in the range of viewing angle $0<\theta_v<1/\gamma$. Since the chance of $\theta_v=0$ is negligibly small, one does not expect purely circularly polarized emission for relativistic ECME. Since the line of sight probability $p(\theta_v) d \theta_v = \sin\theta_v d\theta_v$, one expects that most viewing angles carry significant linear polarization.
Outside the $1/\gamma$ cone, the radiation intensity drops rapidly, the chance of observing an object in these viewing angles diminishes.

}


\begin{thebibliography}{}
\expandafter\ifx\csname natexlab\endcsname\relax\def\natexlab#1{#1}\fi
\providecommand{\url}[1]{\href{#1}{#1}}
\providecommand{\dodoi}[1]{doi:~\href{http://doi.org/#1}{\nolinkurl{#1}}}
\providecommand{\doeprint}[1]{\href{http://ascl.net/#1}{\nolinkurl{http://ascl.net/#1}}}
\providecommand{\doarXiv}[1]{\href{https://arxiv.org/abs/#1}{\nolinkurl{https://arxiv.org/abs/#1}}}

\bibitem[{{Barrett} {et~al.}(2020){Barrett}, {Dieck}, {Beasley}, {Mason}, \& {Singh}}]{Barrett2020}
{Barrett}, P., {Dieck}, C., {Beasley}, A.~J., {Mason}, P.~A., \& {Singh}, K.~P. 2020, Advances in Space Research, 66, 1226, \dodoi{10.1016/j.asr.2020.04.007}

\bibitem[{{Bilbao} {et~al.}(2024){Bilbao}, {Silva}, \& {Silva}}]{Bilbao2024}
{Bilbao}, P.~J., {Silva}, T., \& {Silva}, L.~O. 2024, arXiv e-prints, arXiv:2409.18955, \dodoi{10.48550/arXiv.2409.18955}

\bibitem[{{Buckley} {et~al.}(2017){Buckley}, {Meintjes}, {Potter}, {Marsh}, \& {G{\"a}nsicke}}]{Buckley2017}
{Buckley}, D.~A.~H., {Meintjes}, P.~J., {Potter}, S.~B., {Marsh}, T.~R., \& {G{\"a}nsicke}, B.~T. 2017, Nature Astronomy, 1, 0029, \dodoi{10.1038/s41550-016-0029}

\bibitem[{{Caleb} {et~al.}(2024){Caleb}, {Lenc}, {Kaplan}, {Murphy}, {Men}, {Shannon}, {Ferrario}, {Rajwade}, {Clarke}, {Giacintucci}, {Hurley-Walker}, {Hyman}, {Lower}, {McSweeney}, {Ravi}, {Barr}, {Buchner}, {Flynn}, {Hessels}, {Kramer}, {Pritchard}, \& {Stappers}}]{Caleb2024}
{Caleb}, M., {Lenc}, E., {Kaplan}, D.~L., {et~al.} 2024, Nature Astronomy, \dodoi{10.1038/s41550-024-02277-w}

\bibitem[{{Chanmugam} \& {Dulk}(1982)}]{Chanmugam&Dulk1982}
{Chanmugam}, G., \& {Dulk}, G.~A. 1982, \apjl, 255, L107, \dodoi{10.1086/183779}

\bibitem[{{Cooper} \& {Wadiasingh}(2024)}]{Cooper2024}
{Cooper}, A.~J., \& {Wadiasingh}, Z. 2024, \mnras, 533, 2133, \dodoi{10.1093/mnras/stae1813}

\bibitem[{{Dall'Osso} {et~al.}(2006){Dall'Osso}, {Israel}, \& {Stella}}]{DallOsso2006}
{Dall'Osso}, S., {Israel}, G.~L., \& {Stella}, L. 2006, \aap, 447, 785, \dodoi{10.1051/0004-6361:20052843}

\bibitem[{{de Ruiter} {et~al.}(2024){de Ruiter}, {Rajwade}, {Bassa}, {Rowlinson}, {Wijers}, {Kilpatrick}, {Stefansson}, {Callingham}, {Hessels}, {Clarke}, {Peters}, {Wijnands}, {Shimwell}, {ter Veen}, {Morello}, {Zeimann}, \& {Mahadevan}}]{Ruiter2024}
{de Ruiter}, I., {Rajwade}, K.~M., {Bassa}, C.~G., {et~al.} 2024, arXiv e-prints, arXiv:2408.11536, \dodoi{10.48550/arXiv.2408.11536}

\bibitem[{{Dulk} {et~al.}(1983){Dulk}, {Bastian}, \& {Chanmugam}}]{Dulk1983}
{Dulk}, G.~A., {Bastian}, T.~S., \& {Chanmugam}, G. 1983, \apj, 273, 249, \dodoi{10.1086/161363}

\bibitem[{{Geng} {et~al.}(2016){Geng}, {Zhang}, \& {Huang}}]{Geng2016}
{Geng}, J.-J., {Zhang}, B., \& {Huang}, Y.-F. 2016, \apjl, 831, L10, \dodoi{10.3847/2041-8205/831/1/L10}

\bibitem[{{Goldreich} \& {Julian}(1969)}]{Goldreich&Julian1969}
{Goldreich}, P., \& {Julian}, W.~H. 1969, \apj, 157, 869, \dodoi{10.1086/150119}

\bibitem[{{Goldreich} \& {Lynden-Bell}(1969)}]{Goldreich&Lynden-Bell1969}
{Goldreich}, P., \& {Lynden-Bell}, D. 1969, \apj, 156, 59, \dodoi{10.1086/149947}

\bibitem[{{Hansen} \& {Lyutikov}(2001)}]{Hansen&Lyutikov2001}
{Hansen}, B. M.~S., \& {Lyutikov}, M. 2001, \mnras, 322, 695, \dodoi{10.1046/j.1365-8711.2001.04103.x}

\bibitem[{{Hurley-Walker} {et~al.}(2022){Hurley-Walker}, {Zhang}, {Bahramian}, {McSweeney}, {O'Doherty}, {Hancock}, {Morgan}, {Anderson}, {Heald}, \& {Galvin}}]{Hurley-Walker2022}
{Hurley-Walker}, N., {Zhang}, X., {Bahramian}, A., {et~al.} 2022, \nat, 601, 526, \dodoi{10.1038/s41586-021-04272-x}

\bibitem[{{Hurley-Walker} {et~al.}(2023){Hurley-Walker}, {Rea}, {McSweeney}, {Meyers}, {Lenc}, {Heywood}, {Hyman}, {Men}, {Clarke}, {Coti Zelati}, {Price}, {Horv{\'a}th}, {Galvin}, {Anderson}, {Bahramian}, {Barr}, {Bhat}, {Caleb}, {Dall'Ora}, {de Martino}, {Giacintucci}, {Morgan}, {Rajwade}, {Stappers}, \& {Williams}}]{Hurley-Walker2023}
{Hurley-Walker}, N., {Rea}, N., {McSweeney}, S.~J., {et~al.} 2023, \nat, 619, 487, \dodoi{10.1038/s41586-023-06202-5}

\bibitem[{{Hurley-Walker} {et~al.}(2024){Hurley-Walker}, {McSweeney}, {Bahramian}, {Rea}, {Horvath}, {Buchner}, {Williams}, {Meyers}, {Strader}, {Aydi}, {Urquhart}, {Chomiuk}, {Galvin}, {Coti Zelati}, \& {Bailes}}]{Hurley-Walker2024}
{Hurley-Walker}, N., {McSweeney}, S.~J., {Bahramian}, A., {et~al.} 2024, arXiv e-prints, arXiv:2408.15757, \dodoi{10.48550/arXiv.2408.15757}

\bibitem[{{Hyman} {et~al.}(2005){Hyman}, {Lazio}, {Kassim}, {Ray}, {Markwardt}, \& {Yusef-Zadeh}}]{Hyman2005}
{Hyman}, S.~D., {Lazio}, T. J.~W., {Kassim}, N.~E., {et~al.} 2005, \nat, 434, 50, \dodoi{10.1038/nature03400}

\bibitem[{{Kaplan} {et~al.}(2008){Kaplan}, {Hyman}, {Roy}, {Bandyopadhyay}, {Chakrabarty}, {Kassim}, {Lazio}, \& {Ray}}]{Kaplan2008}
{Kaplan}, D.~L., {Hyman}, S.~D., {Roy}, S., {et~al.} 2008, \apj, 687, 262, \dodoi{10.1086/591436}

\bibitem[{{Katz}(2022)}]{katz22}
{Katz}, J.~I. 2022, \apss, 367, 108, \dodoi{10.1007/s10509-022-04146-2}

\bibitem[{{Lai}(2012)}]{Lai2012}
{Lai}, D. 2012, \apjl, 757, L3, \dodoi{10.1088/2041-8205/757/1/L3}

\bibitem[{{Laine} \& {Lin}(2012)}]{Laine2012}
{Laine}, R.~O., \& {Lin}, D. N.~C. 2012, \apj, 745, 2, \dodoi{10.1088/0004-637X/745/1/2}

\bibitem[{Lee {et~al.}(2025)Lee, Caleb, Murphy, Lenc, Kaplan, Ferrario, Wadiasingh, Anumarlapudi, Hurley-Walker, Karambelkar, Ocker, McSweeney, Qiu, Rajwade, Zic, Bannister, Bhat, Deller, Dobie, Driessen, Gendreau, Glowacki, Gupta, Jahns-Schindler, Jaini, James, Kasliwal, Lower, Shannon, Uttarkar, Wang, \& Wang}]{Lee2025}
Lee, Y. W.~J., Caleb, M., Murphy, T., {et~al.} 2025, Nature Astronomy, \dodoi{10.1038/s41550-024-02452-z}

\bibitem[{{Li} {et~al.}(2024){Li}, {Yuan}, {Wu}, {Yan}, {Lv}, {Tsai}, {Wang}, {Zhu}, {Deng}, {Lan}, {Xu}, {Chen}, {Meng}, {Li}, {Li}, {Zhou}, {Yang}, {Xue}, {Lu}, {Miao}, {Wang}, {Niu}, {Fang}, {Fu}, {Feng}, {Zhang}, {Jiang}, {Miao}, {Chen}, {Sun}, {Yang}, {Deng}, {Dai}, {Chen}, {Yao}, {Liu}, {Li}, {Zhang}, {Yang}, {Zhou}, {Yiyizhou}, {Zhang}, {Niu}, {Zhao}, {Zhang}, {Peng}, {Wu}, \& {Wang}}]{LiD2024}
{Li}, D., {Yuan}, M., {Wu}, L., {et~al.} 2024, arXiv e-prints, arXiv:2411.15739, \dodoi{10.48550/arXiv.2411.15739}

\bibitem[{{Marsh} {et~al.}(2016){Marsh}, {G{\"a}nsicke}, {H{\"u}mmerich}, {Hambsch}, {Bernhard}, {Lloyd}, {Breedt}, {Stanway}, {Steeghs}, {Parsons}, {Toloza}, {Schreiber}, {Jonker}, {van Roestel}, {Kupfer}, {Pala}, {Dhillon}, {Hardy}, {Littlefair}, {Aungwerojwit}, {Arjyotha}, {Koester}, {Bochinski}, {Haswell}, {Frank}, \& {Wheatley}}]{Marsh2016}
{Marsh}, T.~R., {G{\"a}nsicke}, B.~T., {H{\"u}mmerich}, S., {et~al.} 2016, \nat, 537, 374, \dodoi{10.1038/nature18620}

\bibitem[{{McWilliams} \& {Levin}(2011)}]{McWilliams2011}
{McWilliams}, S.~T., \& {Levin}, J. 2011, \apj, 742, 90, \dodoi{10.1088/0004-637X/742/2/90}

\bibitem[{{Melrose}(2017)}]{Melrose2017}
{Melrose}, D.~B. 2017, Reviews of Modern Plasma Physics, 1, 5, \dodoi{10.1007/s41614-017-0007-0}

\bibitem[{{Men} {et~al.}(2025){Men}, {McSweeney}, {Hurley-Walker}, {Barr}, \& {Stappers}}]{Men2025}
{Men}, Y., {McSweeney}, S., {Hurley-Walker}, N., {Barr}, E., \& {Stappers}, B. 2025, arXiv e-prints, arXiv:2501.10528, \dodoi{10.48550/arXiv.2501.10528}

\bibitem[{{Papitto} {et~al.}(2013){Papitto}, {Ferrigno}, {Bozzo}, {Rea}, {Pavan}, {Burderi}, {Burgay}, {Campana}, {di Salvo}, {Falanga}, {Filipovi{\'c}}, {Freire}, {Hessels}, {Possenti}, {Ransom}, {Riggio}, {Romano}, {Sarkissian}, {Stairs}, {Stella}, {Torres}, {Wieringa}, \& {Wong}}]{papitto13}
{Papitto}, A., {Ferrigno}, C., {Bozzo}, E., {et~al.} 2013, \nat, 501, 517, \dodoi{10.1038/nature12470}

\bibitem[{{Piro}(2012)}]{Piro2012}
{Piro}, A.~L. 2012, \apj, 755, 80, \dodoi{10.1088/0004-637X/755/1/80}

\bibitem[{{Rea} {et~al.}(2022){Rea}, {Coti Zelati}, {Dehman}, {Hurley-Walker}, {de Martino}, {Bahramian}, {Buckley}, {Brink}, {Kawka}, {Pons}, {Vigan{\`o}}, {Graber}, {Ronchi}, {Pardo Araujo}, {Borghese}, {Parent}, \& {Galvin}}]{Rea2022}
{Rea}, N., {Coti Zelati}, F., {Dehman}, C., {et~al.} 2022, \apj, 940, 72, \dodoi{10.3847/1538-4357/ac97ea}

\bibitem[{{Rea} {et~al.}(2024){Rea}, {Hurley-Walker}, {Pardo-Araujo}, {Ronchi}, {Graber}, {Coti Zelati}, {de Martino}, {Bahramian}, {McSweeney}, {Galvin}, {Hyman}, \& {Dall'Ora}}]{Rea2024}
{Rea}, N., {Hurley-Walker}, N., {Pardo-Araujo}, C., {et~al.} 2024, \apj, 961, 214, \dodoi{10.3847/1538-4357/ad165d}

\bibitem[{{Ruderman} \& {Sutherland}(1975)}]{Ruderman1975}
{Ruderman}, M.~A., \& {Sutherland}, P.~G. 1975, \apj, 196, 51, \dodoi{10.1086/153393}

\bibitem[{{Stanway} {et~al.}(2018){Stanway}, {Marsh}, {Chote}, {G{\"a}nsicke}, {Steeghs}, \& {Wheatley}}]{Stanway2018}
{Stanway}, E.~R., {Marsh}, T.~R., {Chote}, P., {et~al.} 2018, \aap, 611, A66, \dodoi{10.1051/0004-6361/201732380}

\bibitem[{{Twiss}(1958)}]{Twiss1958}
{Twiss}, R.~Q. 1958, Australian Journal of Physics, 11, 564, \dodoi{10.1071/PH580564}

\bibitem[{{Wadiasingh} {et~al.}(2020){Wadiasingh}, {Beniamini}, {Timokhin}, {Baring}, {van der Horst}, {Harding}, \& {Kazanas}}]{Wadiasingh20}
{Wadiasingh}, Z., {Beniamini}, P., {Timokhin}, A., {et~al.} 2020, \apj, 891, 82, \dodoi{10.3847/1538-4357/ab6d69}

\bibitem[{{Wang} {et~al.}(2016){Wang}, {Yang}, {Wu}, {Dai}, \& {Wang}}]{WangJS2016}
{Wang}, J.-S., {Yang}, Y.-P., {Wu}, X.-F., {Dai}, Z.-G., \& {Wang}, F.-Y. 2016, \apjl, 822, L7, \dodoi{10.3847/2041-8205/822/1/L7}

\bibitem[{{Wang} {et~al.}(2024){Wang}, {Rea}, {Bao}, {Kaplan}, {Lenc}, {Wadiasingh}, {Hare}, {Zic}, {Anumarlapudi}, {Bera}, {Beniamini}, {Cooper}, {Clarke}, {Deller}, {Dawson}, {Glowacki}, {Hurley-Walker}, {McSweeney}, {Polisensky}, {Peters}, {Younes}, {Bannister}, {Caleb}, {Dage}, {James}, {Kasliwal}, {Karambelkar}, {Lower}, {Mori}, {Ocker}, {P{\'e}rez-Torres}, {Qiu}, {Rose}, {Shannon}, {Taub}, {Wang}, {Wang}, {Zhao}, {Bhat}, {Dobie}, {Driessen}, {Murphy}, {Jaini}, {Deng}, {Jahns-Schindler}, {Lee}, {Pritchard}, {Tuthill}, \& {Thyagarajan}}]{WangZT2024}
{Wang}, Z., {Rea}, N., {Bao}, T., {et~al.} 2024, arXiv e-prints, arXiv:2411.16606, \dodoi{10.48550/arXiv.2411.16606}

\bibitem[{{Warner}(1995)}]{warner95}
{Warner}, B. 1995, {Cataclysmic variable stars}, Vol.~28

\bibitem[{{Wu} \& {Lee}(1979)}]{Wu&Lee1979}
{Wu}, C.~S., \& {Lee}, L.~C. 1979, \apj, 230, 621, \dodoi{10.1086/157120}

\bibitem[{{Wu} {et~al.}(2002){Wu}, {Cropper}, {Ramsay}, \& {Sekiguchi}}]{Wu2002}
{Wu}, K., {Cropper}, M., {Ramsay}, G., \& {Sekiguchi}, K. 2002, \mnras, 331, 221, \dodoi{10.1046/j.1365-8711.2002.05190.x}

\bibitem[{{Wu} \& {Wickramasinghe}(1993)}]{Wu&Wickramasinghe1993}
{Wu}, K., \& {Wickramasinghe}, D.~T. 1993, in Cataclysmic Variables and Related Physics, ed. O.~{Regev} \& G.~{Shaviv}, Vol.~10, 336

\bibitem[{{Zhang}(2023)}]{Zhang2023RMP}
{Zhang}, B. 2023, Reviews of Modern Physics, 95, 035005, \dodoi{10.1103/RevModPhys.95.035005}

\bibitem[{{Zhang} \& {Gil}(2005)}]{ZhangGil2005}
{Zhang}, B., \& {Gil}, J. 2005, \apjl, 631, L143, \dodoi{10.1086/497428}

\bibitem[{{Zhu} \& {Xu}(2006)}]{Zhu&Xu2006}
{Zhu}, W.~W., \& {Xu}, R.~X. 2006, \mnras, 365, L16, \dodoi{10.1111/j.1745-3933.2005.00117.x}

\end{thebibliography}

\end{document}